\documentclass[pre,twocolumn,floatfix]{revtex4}

\usepackage{graphicx}
\usepackage{dcolumn}
\usepackage{bm}

\usepackage{amssymb}     
\usepackage{amsmath}
\begin{document}


\title{Ratchet behavior in nonlinear Klein-Gordon systems with point-like inhomogeneities}

\author{Luis Morales--Molina}
\email{Luis.Morales-Molina@uni-bayreuth.de}
\author{Franz G.\ Mertens}
\email{Franz.Mertens@uni-bayreuth.de}
\affiliation{Physikalisches Institut, Universit\"at Bayreuth, D-85440 Bayreuth,
Germany}

\author{Angel S\'anchez}%
\homepage{http://gisc.uc3m.es/~anxo}
\affiliation{%
Grupo Interdisciplinar de Sistemas Complejos (GISC) and
Departamento de Matem\'aticas,
Universidad Carlos III de Madrid, Avenida de la Universidad 30, 28911
Legan\'es, Madrid, Spain}%
\affiliation{%
Instituto de Biocomputaci\'on y F\'\i sica de Sistemas Complejos,
Universidad de Zaragoza, 50009 Zaragoza, Spain
}%

\date{\today}

\begin{abstract}
We investigate the ratchet dynamics of nonlinear Klein-Gordon kinks in a periodic, asymmetric lattice of point-like inhomogeneities. We explain the underlying
rectification mechanism within a collective coordinate framework,
which shows that such system behaves as a rocking ratchet for point
particles.
Careful attention is given to the kink width dynamics and its role in the transport. 
We also analyze the robustness of our kink rocking ratchet in the presence of noise. 
We show that the noise activates unidirectional motion in a parameter range where such motion is not observed in the noiseless case. This is 
subsequently corroborated by the collective variable theory.
An explanation for this new phenomenom is given. 

\end{abstract}


\maketitle

\section{Introduction}
The study of transport mechanisms at the mesoscale level is of great importance nowadays. Specifically,
the so-called ratchet systems have shown to be proper candidates for explaining unidirectional motion to biological systems \cite{Maddox}, and have important physical applications for nano- and micro-scale
technologies \cite{superconductor,Linke,Grifoni}.
Many of these models have been developed in the simple picture of point-like
particles \cite{hangi,hangi2,sironis} (see the reviews \cite{reiman,hangi3}
for details). Such scenario has been subsequently generalized to spatially
extended systems \cite{extended,falo,falo2}, where much attention has
been paid to situations where the net motion arises through time-symmetry breaking
\cite{flach-Sal,Niur}. This kind of ratchet phenomenon has been recently observed in
long Josephson junction (LJJ) devices \cite{Ustinov}.
Another possibility that has been considered in the literature is that of nonlinear Klein-Gordon system
where the on-site potential is ratchet-like \cite{salerno2}. Notwithstanding, 
to our knowledge the case of spatial-symmetry breaking by inhomogeneities has not
been studied in depth. One such study has been done by Carapella et al. who used an inhomogeneous magnetic field to create an effective inhomogeneous junction profile for fluxons to propagate \cite{Carapella}.
Recently an alternative to the generation of motion for extended systems with a disorder in the chain has been proposed \cite{pub}.
The novelty of the procedure is the design of a ratchet device from 
a lattice of {\em point-like inhomogeneities}.
For this system, net motion arises from the
interplay between disorder and nonlinearity of the nonlinear systems \cite{angel0}.

In this paper we elaborate on the preliminary results reported in \cite{pub}. 
Our aim is to carry out an in-depth analysis of the system, including a careful 
comparison to related point-like ratchets \cite{floria} and an extension of 
our results, originally obtained for the sine-Gordon (sG) model, to other 
nonlinear Klein-Gordon models such as the $\phi^4$ equation.
Additional motivation for
this work arises from research 
on models of energy propagation along microtubule filaments
inside the cells \cite{Sataric}. 
This application is specially interesting in view of 
the possible connection with the dynamics of transport in molecular motors
in biological systems, with features similar to those of solitons as extended objects. 
In this context, the present work sheds light on the role played by the length scale
competition between the point-like inhomogeneities (disorder) and the size of kinks
in the transport dynamics. For this purpose, we use the framework of collective
coordinates (CC) in order to gain insight in the cause of the motion and the degrees of freedom that take part in it. Emphasis will be given to the kink width oscillations and their role in the transport properties: Indeed, 
in general, the width of the nonlinear topological excitations is crucial for
the movement of these coherent excitations. The coupling between the translational
and kink width degrees is such that motion takes place \cite{Niur}. In particular,
in the present work we will see that the oscillation range of the kink width
is determined by the interplay with the inhomogeneities.
An additional, relevant
issue is the analysis of the motion dynamics under thermal fluctuations.
In ratchets, the noise is an important source of energy and,
for some biological systems, it is regarded as the main cause of transport.    
Here we will consider the robustness of our rocking ratchet system under
thermal fluctuations. In this case, activation of unidirectional motion was observed for a certain range of frequencies in the simulations as well as in the CC.

In order to achieve the above mentioned goals, our paper is organized as follows:
In Sec.\ II we formulate the basis for the ratchet device and explain
the origin and physical reasons of the rectification process.
A discussion in the CC framework is devoted to the length scale competition
between the inhomogeneities and the kink width,
and its influence on the motion dynamics.
In the same context, we establish an analogy with a simple model used for
describing unidirectional motion in molecular motors \cite{floria}, pointing 
out the relevant role of the kink width for the ratchet dynamics and its 
application in biological systems. Subsequently,
an analysis of the efficiency in terms of the quantization of the transport is done for
the relevant parameters of our ratchet system, including the interference
effects among the inhomogeneities. An example is given for the $\phi^4$ model,
not only for its known rich internal dynamics reinforced by the presence of
an internal mode, but also for its potential application to macromolecules 
(e.g., in transfer of energy in microtubules \cite{Sataric}).  
Next, in Sec.\ III, we analyze 
the kink dynamics subject to noise. In particular, a new phenomenon of
activation of motion induced by noise is described.
In order to explain this new behavior,
different collective coordinate approaches are implemented and thoroughly 
discussed.
Finally, in the last Section we summarize the main contributions of our work and
make a discussion about the perspectives on this line. We include 
appendixes where we detail the CC approaches for one and two collective
variables, extending the {\em Generalized Traveling Wave Ansatz} (GTWA) to the case
where inhomogeneities and noise act together with damping and ac forces.

\section{Ratchet model and transport}

\subsection{Model} 

Kink dynamics in the presence of inhomogeneities can exhibit 
different and interesting behaviors, depending on the interplay between those
inhomogeneities and the nonlinearity \cite{kivshar,angel} among other factors. 
The generation of net motion using a lattice of point-like inhomogeneities 
is a good example of such non-trivial phenomenon \cite{pub}. 
Although in that previous work the problem was discussed for the sG model,
it can be generalized in principle for any nonlinear Klein Gordon system.
Therefore, aiming at that general viewpoint,
for our analysis we formulate the model in a general way  as follows:
\begin{equation}
\ \phi_{tt}+\beta\phi_{t}-\phi_{xx}+ \frac{\partial\widetilde{ U}}{\partial\phi }[1+V(x)]=A\sin(\omega t+\delta_{0}),  \label{1}
\end{equation}
where $\widetilde{U}(\phi)$ is the potential for the nonlinear Klein-Gordon equations and $ A\sin(\omega t+\delta_{0})\equiv f(t)$ is an external ac force whose parameters $A$, $\omega$ and $\delta_{0}$ represent the amplitude, frequency and phase of the periodic force, respectively. 
In particular, we will focus on the $\phi^4$ and sG models as specific 
examples; the corresponding nonlinear potentials are
$\widetilde{U}(\phi)=\frac14(\phi^2-1)^2$
and $\widetilde{U}(\phi)=[1-\cos(\phi)]$ respectively.

For our potential to exhibit ratchet-like phenomenon, we choose 
$V(x)$ to be given by a periodically repeated unit cell,
formed by an asymmetric array of delta functions (inhomogeneities).
The unit cell configuration, of length $L$, 
is defined by three inhomogeneities, the first one located at the 
beginning of the cell, the second at distance $a$ from the first, and
the third at distance $b$ from the second. The corresponding mathematical
expression is 
\begin{eqnarray}\label{2}
V(x)=\epsilon \sum_{n}\left[\delta(x-x_{1}-nL)+\delta(x-x_{2}-nL)\right.\nonumber\\
\left.+\delta(x-x_{3}-nL)\right]
\end{eqnarray}
where the parameters satisfy the following constraints  $a,b,c\sim l_{0}$ (static kink width in absence of inhomogeneities);  $a,b<c$  with $a\neq b$, 
where  $L=a+b+c$, $a=x_{2}-x_{1}$, $b=x_{3}-x_{2}$ and $c=L+x_{1}-x_{3}$, with $x_{1}<x_{2}<x_{3}$.
For our study we have taken $\epsilon>0$, where
in the case of sG, specifically for LJJ the point-like inhomogeneities represent
microshorts \cite{mclaughlin,malomed}. However, the same scheme of arrays of point like inhomogeneities can be implemented for $\epsilon <0$, where particularly for the sG model many works have been devoted \cite{Ustinov2}.
The choice of three inhomogeneities is inspired by biological polymers like DNA where the existence of three bases per codon seems to be the ideal configuration for the occurrence of net transport \cite{cocho}.
In principle it is possible to obtain unidirectional motion by using an array whose configuration presents more than three inhomogeneities per unit cell 
if the distances between the delta functions are
in the same length scale as the kink width (otherwise, different
behaviors could arise like obtained e.g. in \cite{angel}).
However, the inclusion of more inhomogeneities diminishes the efficiency of
the transport as we will see below.

\begin{figure}
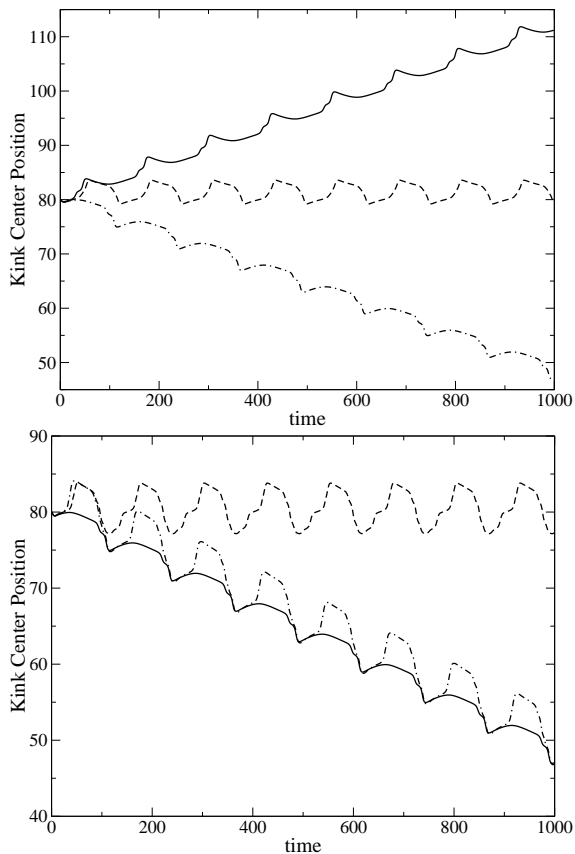

\begin{center}
\includegraphics[width=7.5cm]{0.35-omega=0.05-x1=0.5-x3=2.3-x2=1,1.4.1.8-L=4.eps}

\includegraphics[width=7.5cm]{Figuras-omega=0.05ratchet0.eps}
\end{center}
\caption{Simulations of Eq.(\ref{1}) for sG case: Position of kink center vs time:
Upper panel:
Different arrays with the same amplitude of the force $A=0.35$:  $x_{1}=0.5$,
$x_{2}=1.8$, $x_{3}=2.3$ ($a>b$) (solid line);
$x_{1}=0.5$, $x_{2}=1.4$, $x_{3}=2.3$ ($a=b$) (dashed line);
$x_{1}=0.5$, $x_{2}=1$, $x_{3}=2.3$ ($a<b$) (dash-dotted line).
Lower panel:
For different amplitudes of the ac force: $A=0.35$ (solid line); $A=0.45$
(dashed line); $A=0.50$ (dash-dotted line) with the array $x_{1}=0.5$,
$x_{2}=1$, $x_{3}=2.3$. The other parameters used are
$\omega=0.05$, $\epsilon=0.8$, $\delta_{0}=\pi$ and period $L=4$.}
\label{ratchet0}
\end{figure}

\subsection{Simulations of the model}

Contrary to the case of point particles, where motion through point-like
inhomogeneities (delta functions) is physically meaningless, in our case we deal
with kinks (extended objects) with a determined width. This is an intrinsic feature of
these nonlinear excitations, and correspondingly the competition between their width
and the distances among the inhomogeneities is crucial for the kink motion.
The interference effects among the inhomogeneities \cite{angel2} create an
effective potential for the motion of the kink center, the location of the
inhomogeneities determining the direction of motion.
For the particular configuration of three inhomogeneities per unit cell,
directional motion takes place only under the condition $a\neq b $.
In the upper panel of Fig.\ \ref{ratchet0}, results of simulations of Eq.\ (\ref{1})
for the sG case with different values of $a$ and $b$
are depicted. Such a picture shows that our ratchet device is an authentic rectifier. 
In addition,
as in standard ratchet systems, the directional motion of the kink center takes
place only for certain values of the amplitude of the ac force (see lower panel 
of Fig.\ \ref{ratchet0}), a behavior that is dependent on the ac force frequency.  
A more detailed picture of the dynamics of the mean velocity for the sG kink center
as a function of the ac force amplitude for different frequencies can be found
in \cite{pub}.

We have restricted ourselves to the overdamped case by taking $\beta=1$,
where the inertial effects are small, reducing the generation and propagation
of phonons, and for which the kink center moves on a tilted effective potential
due to the external ac force. 
This regime prevents also the dependence on the initial conditions for the
dynamics \cite{Jung}. 
For the integration of Eq.\ (\ref{1}) we have used a Strauss-V\'azquez numerical scheme
\cite{strauss} with free boundary conditions and spatial and temporal steps
$\Delta x=0.1$ and $\Delta t=0.01$, respectively. We have checked our results
with different spatial steps $\Delta x=0.05$ and $\Delta x=0.02$.
The simulations were done for the  spatial interval $[-30,150]$ with
inhomogeneities arranged periodically according to our period in $[0,120]$. 
We have used the numerical representation for delta functions given by
\begin{equation}
\label{deltas}
\delta(x-x_c)\rightarrow \left\{ \begin{array}{c}
\hspace {0.5cm} 1/\Delta x , \hspace {1cm} \vert x-x_c\vert<\Delta x/2\\
 0, \hspace {1cm} \mbox{otherwise.}
\end{array}\right.
\end{equation}
Such representation is not unique but is one of the simplest forms for the numerical implementation \cite{kivshar,angel2}.

\subsection{Collective coordinate approach}

In order to understand the dynamics of the kink center different CC approaches have been implemented \cite{pub}.
In this previous work, it was shown that a 1-CC approach is not enough for the correct quantitative description of the results of the simulations even considering possible relativistic effects for high values of ac the force.
However, this approach allows us to understand the underlying physics of this ratchet system in the simplest way.  
On the other hand,
satisfactory agreement with the simulations was found by an improved, 2-CC approach,
which takes into account the kink width dynamics \cite{pub}.
At this point we will proceed directly with the discussion of the improved theory.
The corresponding
collective variable equations of two degrees of freedom for the Eq.\ (\ref{1})
(see Appendix B, Eqs.\ (\ref{B21})-(\ref{B22}) for $D=0$) can be expressed as
\begin{equation}
\ M_{0}l_{0}\frac{\ddot{X}}{l}+\beta M_{0}l_{0}\frac{\dot{X}}{l}-M_{0}l_{0}
\frac{\dot{X}\dot{l}}{l^2}=-\frac{\partial U}{\partial X}-qf(t),\label{4}
\end{equation}
\begin{eqnarray}
\alpha
  M_{0}l_{0}\frac{\ddot{l}}{l}+\beta\alpha
  M_{0}l_{0}\frac{\dot{l}}{l}+\frac{1}{2}M_{0}l_{0}\frac{\dot{X}^2}{l^2}-\frac{1}{2}  \alpha M_{0}l_{0}\frac{\dot{l}^{2}}{l^2} 
\nonumber
\\
 =- \frac{\partial U^{int}}{\partial l}-\frac{\partial U}{\partial l},\label{5}
\end{eqnarray}
where $X$ stands for the position of the kink center, $l$ represents the kink 
width, 
the internal potential energy of the kink is     
\begin{equation}
U^{int}=\frac{1}{2} M_{0}\left(\frac{l_{0}}{l}+\frac{l}{l_{0}}\right),\label{6}
\end{equation}
and $U(X,l)$ is an effective potential depending on the specific equation one 
is dealing with. 
In particular, for the sG case $ M_{0}=8$, $l_{0}=1$, $\alpha=\pi^2/12$, $q=2\pi$ 
and the effective potential is given by 
\begin{eqnarray}\label{7}
U(X,l)=2\epsilon \sum_{n}\left[\frac{1}{\cosh^2[(X-x_{1}-nL)/l]}\hspace{2cm}\right. \nonumber \\
\left.+\frac{1}{\cosh^2[(X-x_{2}-nL)/l]}+\frac{1}{\cosh^2[(X-x_{3}-nL)/l]}\right].
\end{eqnarray}

As we can see from the previous equations the kink width dynamics is coupled to
the motion of the center of the kink. Therefore, changes in the kink width directly
affect the translational motion. It is possible to observe, for instance, 
that decreasing the kink width decreases the effective ac force, making necessary
to increase the amplitude of the ac force in order to compensate for such effect.
This is an important factor that explains in part the shift observed in the
locations of the windows of motion of the simulations with respect
to those obtained from the 1-CC approach (see Fig.\ 1b in \cite{pub}). 
Another relevant conclusion is the feedback between 
the effective potential landscape and the kink width, determined in turn by 
the potential. In this fashion, the 2-CC approach reflects the non-trivial 
interaction of the kink with the inhomogeneities, which is otherwise known to exhibit
many counterintuitive phenomena \cite{angel2}.

To deepen our
understanding of the dynamics, let us look into the oscillations of the kink width.
As in the case of simulations we restrict ourselves to the overdamped case
(taking $\beta=1$). 
\begin{figure}
\begin{center}
\includegraphics[width=8cm]{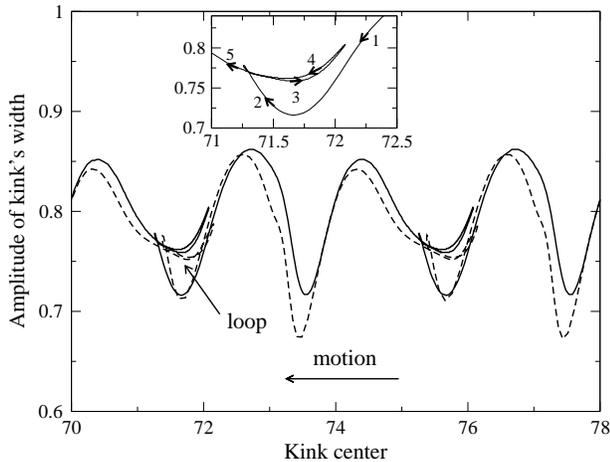}
\end{center}
\caption{sG: Amplitude of the kink width versus kink center. Simulation (solid line);
2-CC Approach Eqs.\ (\ref{4}-\ref{5}) (dashed line). The parameters are
$\omega=0.1$, $A=0.44$, for the array $x_{1}=0.5$, $x_{2}=1$, $x_{3}=2.3$,
period $L=4$ and $\epsilon=0.8$. See text for a discussion of the loop.
Inset: enlargement of the loop indicated by an arrow in the main figure. The motion of the kink center is indicated by numbered arrows.}
\label{ancho}
\end{figure}
\begin{figure}
\begin{center}
\includegraphics[width=8.3cm]{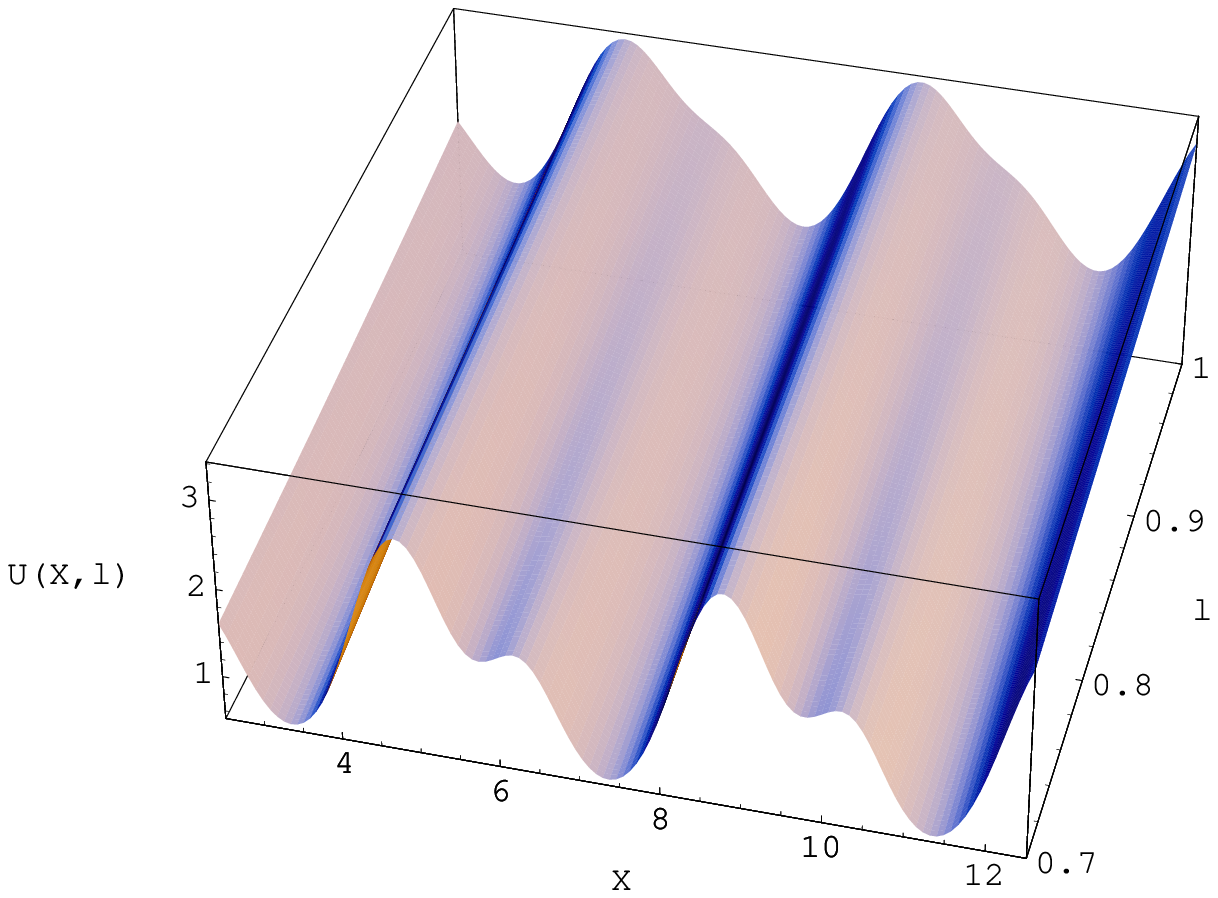}
\includegraphics[width=8.5cm]{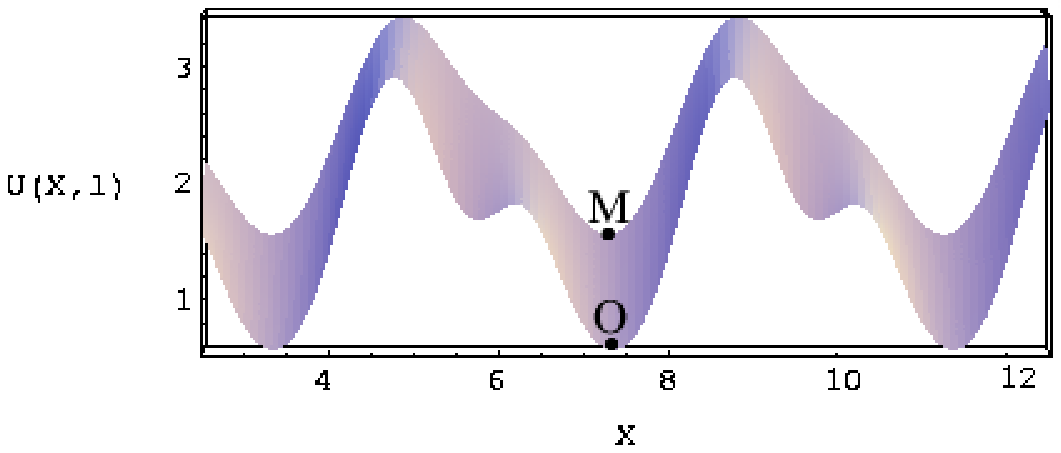}
\end{center}
\caption{a) 3D plot of the effective potential Eq.\ (\ref{7}).
b) Frontal view rotated by a small angle.
The array used is the same as in Fig.\ref{ancho}.}
\label{rat3pot}
\end{figure} 
\begin{figure}
\begin{center}
\includegraphics[width=8cm]{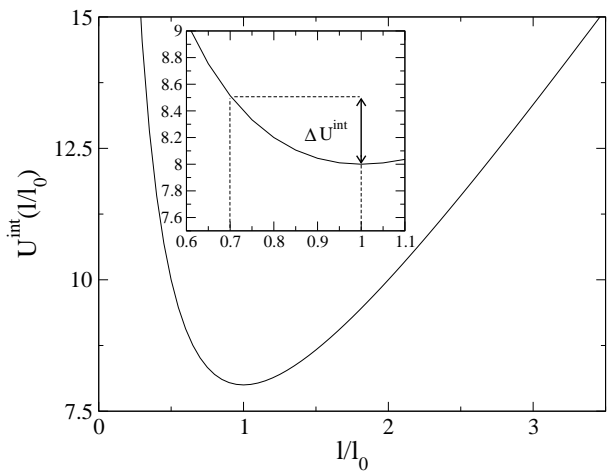}
\end{center}
\caption{Internal potential energy vs normalized kink width, Eq.\ (\ref{6})
(sG case). The inset shows the part of curve where the oscillations of the kink
width take place.}
\label{curveUint}
\end{figure}
A picture of the kink width oscillations versus kink center position is shown in
Fig.\ \ref{ancho}. 
As we can see, the agreement with the CC approach is excellent,
confirming the validity of our predictions. It is particularly interesting 
the existence 
of loops, which arises as a consequence
of the kink center motion back and forth into the wells of the effective potential
(see Fig.\ \ref{rat3pot}), i.e., describing half an oscillation before
overcoming the barrier.

Interestingly,
another feature that stands out clearly is that the oscillations are around a
value different from $l_{0}=1$, the width of the unperturbed kink. Figure \ref{ancho}
shows that they take place around $l\approx 0.8$ and, furthermore,
that $l_{0}$ is not even included in the range of oscillations.
This phenomenon is the result of the balance between two opposite forces. On the
one hand, 
the inclusion of inhomogeneities increases the potential energy of the system.
This fact is reflected in the effective potential energy landscape
Fig.\ \ref{rat3pot}. Such picture shows that when the kink width decreases,
the potential energy decreases as well. Taking two points with the same value for
$X$ but with different kink widths $l$, for example  $M$ and $O$ in Fig.\
\ref{rat3pot}b), we see this difference in potential energy, i.e.,
$E_{M}>E_{O}$ where $l_{M}>l_{O}$. Therefore, as the tendency of the system
is to go to the minimum of the potential energy, the kink width would decrease. 
On the other hand, the kink internal potential energy, Eq.\ (\ref{6}), has a
minimum in $l_{0}$, and hence the energy increases when the kink width
decreases (see Fig. (\ref{curveUint}) for $l<l_{0}$); notice that 
the first term of this equation accounts for a
repulsive interaction and the second is given by an attractive one. 
As a result of this balance, a new minimum will appear for the oscillations of
the kink width. It is important to note that the difference
 $\Delta U^{int}$ of the internal potential
energy for the kink width $l=0.7$ with respect to the value $l_{0}=1$
(inset of Fig.\ \ref{curveUint}) is of the same order as the energy difference 
$E_{M}-E_{O}$ between the points mentioned before for the
effective potential introduced by the inhomogeneities, in agreement with 
this discussion.

\subsection{Related point particle models.}

A problem closely related to our 2-CC approach, given by a point particle ratchet
 with two degrees of freedom, has been studied in \cite{floria}. 
This model was designed for describing molecular motor dynamics consisting of two
particles joined by a spring moving in a ratchet potential. The corresponding
equations of motion are given by
\begin{eqnarray}
\dot{u_1}&=&-\frac{\partial V(u_1)}{\partial u_1}-\frac{\partial W(u_2-u_1)}{\partial u_1}+A \sin(\omega t)+\xi_1(t),\qquad\\
\
\dot{u_2}&=&-\frac{\partial V(u_2)}{\partial u_2}-\frac{\partial W(u_2-u_1)}{\partial u_2}+A\sin(\omega t)+\xi_2(t),\qquad
\end{eqnarray}
where $u_{1}$, $u_{2}$ represent the coordinates of the particles, $V$ is a sawtooth potential and $W$ is the internal potential energy.
Here $\xi_i$ with $i=1,2$ are Gaussian white noises.
Ignoring the noise terms and their influence on the net motion,
we see that the change of variables $X=\frac12( u_1+u_2)$ and $l=u_2-u_1$ casts
the system in a similar shape as Eqs.\ (\ref{4}-\ref{5}) in the overdamped case,
for which in good approximation the inertial terms could be neglected. In such a
new context the variables $X$, $l$ can be interpreted as the mass center and
elongation (distance between the particles) respectively, and obviously they
resemble to the variables  mass center and width of the kink in our system.

Notice that in both models we have an asymmetric potential. In our case it is
given by Eq.\ (\ref{7}), which is asymmetric at the CC level 
if the already mentioned conditions
for the distances between the inhomogeneities are satisfied. 
In both systems, there are internal potential energies that characterize their
elastic properties. 
In the model in \cite{floria}, the internal potential is expressed through
a harmonic function (in the original variables):
\begin{equation}
W(u_1,u_2)=\frac12 k[(u_2-u_1)-l_0]^2,
\end{equation}
which in our collective coordinates can be written as
\begin{equation}
W(l)=\frac12 k[l(t)-l_{0}]^2.
\end{equation}
where $k$ is the elasticity constant.

The links between the two models can be made more explicit by using a value for
$l_0$ close to the minimum around which the kink width oscillates in our 
simulations (cf.\ the discussion in the preceding subsection). \\
 It is important to point out that we have defined $l$ as a kink width variable as given in the expression (\ref{B7})
of the appendix.
However, what the quantity $l$ actually means is the distance at which the kink shape  approaches its asymptotic values, measured from the center. This means that $l$ in our notation is half the ``real kink width''. Consequently the ratio
(real kink width)/(period of the effective potential) becomes $2l_{0}/L\approx 0.4$ for which a very interesting dynamics for point particles dynamics has been reported in related 2-particle model \cite{wang}.
\\
This comparison between our model and that in \cite{floria} allows to point out
their main differences as well. It is particularly important that 
in our framework, the internal energy can describe satisfactorily the
repulsive interaction between real molecules where a van-der-Waals like-force prevents
their overlap. This is very close to what occurs in molecular motors: if we take 
again
the motion of kinesin as an example, this molecule has two dimer heads that act 
as `feet', allowing the molecule to `walk' along a microtubule \cite{kinesin}.
The repulsion would
then appear when the two dimer heads are too close. Such a repulsive interaction 
can not be naturally accounted for 
within the model of two particles. 
For solving this problem the authors of \cite{floria} 
resort to fix arbitrary values for $l_0$ which in our case is not necessary. 
Note, however,  that
in spite of the technical differences between both models,
phenomenologically they are very similar: both of them try to understand how 
the motion of molecular motors, which proceeds in steps accompanied by 
deformations (in the case of kinesin, when one step advances in front of
the other) can arise. The common conclusion is that a point particle ratchet
would not be a good model because the second degree of freedom is needed to 
capture the whole mechanism of the motion. The advantage in our approach is 
that this second degree of freedom arises on its own, without a priori 
constructions, as an emergent property of the nonlinear excitation. 
Recent studies \cite{wang,dan} show similar phenomena for the two degrees of
freedom point particle ratchet of \cite{floria} when the ratchet is of 
flashing type. The close relationship of the model of \cite{floria} to 
ours suggests that nonlinear Klein-Gordon models can also exhibit rectification
working as flashing ratchets, an issue we will address in future work. 

\subsection{Length scales and quantization of transport}

It should be clear, from
the results discussed so far,  that in order to obtain a ratchet device for
extended nonlinear systems with topological nonlinear excitations, the
configuration of the inhomogeneities should be designed in such a way
that the distance between the inhomogeneities is of the order of the
kink width. 
However, this picture is somewhat too simple, and as we will see below,
another important factor to take into account is the existence of 
interference effects. 
Naively, one may try to design a similar ratchet system for the $\phi^4$
equation. Considering only the kink width factor, it would seem that enlarging
the sG array by a factor of $\sqrt{2}$ (the ratio between the kink widths in
both models) similar phenomena would be observed. 
\begin{figure}
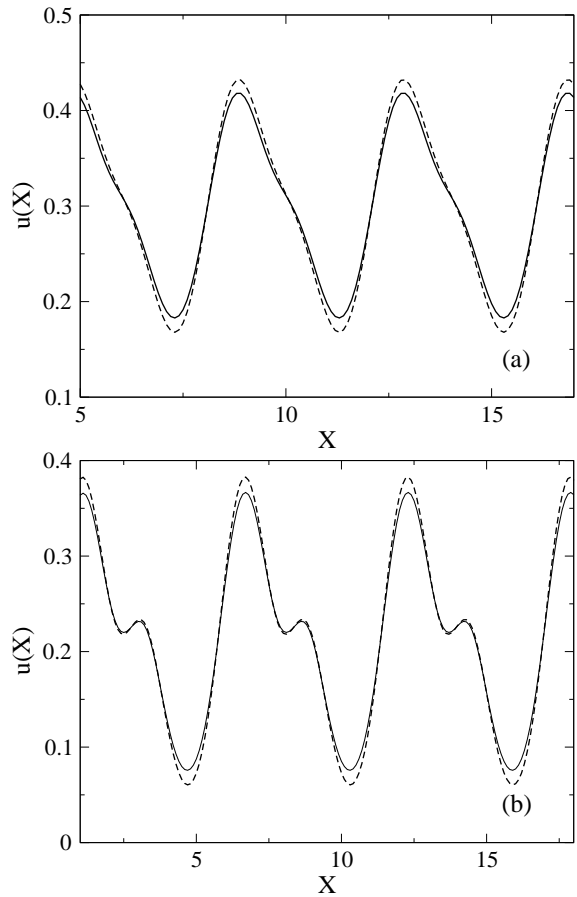

\begin{center} 
\includegraphics[width=7.5cm]{renorpotentL=4-sG-phi4.eps}
\includegraphics[width=7.5cm]{renorpotentL=5.6-sG-phi4.eps}
\end{center}
\caption{ Normalized effective potential for the kink center within the CC approach,
Eq.\ (\ref{13})-(\ref{14}),
for two different arrays with $\epsilon=0.8$. (a): $x_{1}=0.5$,
$x_{2}=1$, $x_{3}=2.3$ and $L=4$. (b): $x_{1}=0.7$, $x_{2}=1.4$, $x_{3}=3.2$, $L=5.6$. In both panels sG (solid line); $\phi^4$ (dashed line).}
\label{ratchet1}
\end{figure} 

Let us make a more specific comparison between both models. To this end,
we use the 1-CC framework in the nonrelativistic approach,
where the equation of motion can be written as
\begin{equation}\label{12}
\ddot{X}+ \beta \dot{X}=-\frac{d
u}{d X}-\frac{q A}{M_{0}}\sin(\omega t +\delta_0),
\end{equation}
where $u=U/M_0$ is the normalized effective potential.
For the sG case we have the following expression 
\begin{equation}\label{13}
u(X)=\frac{2\epsilon}{M_0} \sum_{n}\sum_{i=1}^{3}\frac{1}{\cosh^2[(X-x_{i}-nL)/l_{0}]}
\end{equation}
with $l_{0}=1$, and $M_0=8$, whereas
for the case of $\phi^4$ we have 
\begin{equation}\label{14}
u(X)=\frac{\epsilon}{4 M_0} \sum_{n}\sum_{i=1}^{3}\frac{1}{\cosh^4[(X-x_{i}-nL)/l_{0}]}
\end{equation}
with $l_{0}=\sqrt2$, and $M_0=2\sqrt{2}/3$. The parameter $q$ is the topological charge and it is given by $q=2\pi$ and $q=2$ for the systems sG and $\phi^4$,
respectively. 
The normalized effective potentials for two different arrays are depicted in Fig.\ 
\ref{ratchet1}. Panel a) shows standard asymmetric potentials for ratchet systems
obtained with an array that satisfies the conditions mentioned above for 
the location of the inhomogeneities in the sG case. However, in case b)
the effective potentials obtained for an array approximately given by the
multiplication of the factor $\sqrt2$ of the first one, shows a local minimum
similar to an array of asymmetric double-well traps. This last one has been 
used as a device for generating motion of vortices in superconductor materials
\cite{marchesoni}. 

According to the our previous arguments based on the kink width role,
a similar picture is expected for the normalized effective potential of
$\phi^4$ and sG if the arrays verify the same length ratio as the full 
systems. Strikingly, Fig.\ \ref{ratchet1} shows that the normalized effective
potentials are almost the same {\em but for the same array length}.
This apparent discrepancy can be explained if we take a detailed look at
the potential given by Eqs.\ (\ref{13}-\ref{14}) for both cases (sG and $\phi^4$).
\begin{figure}
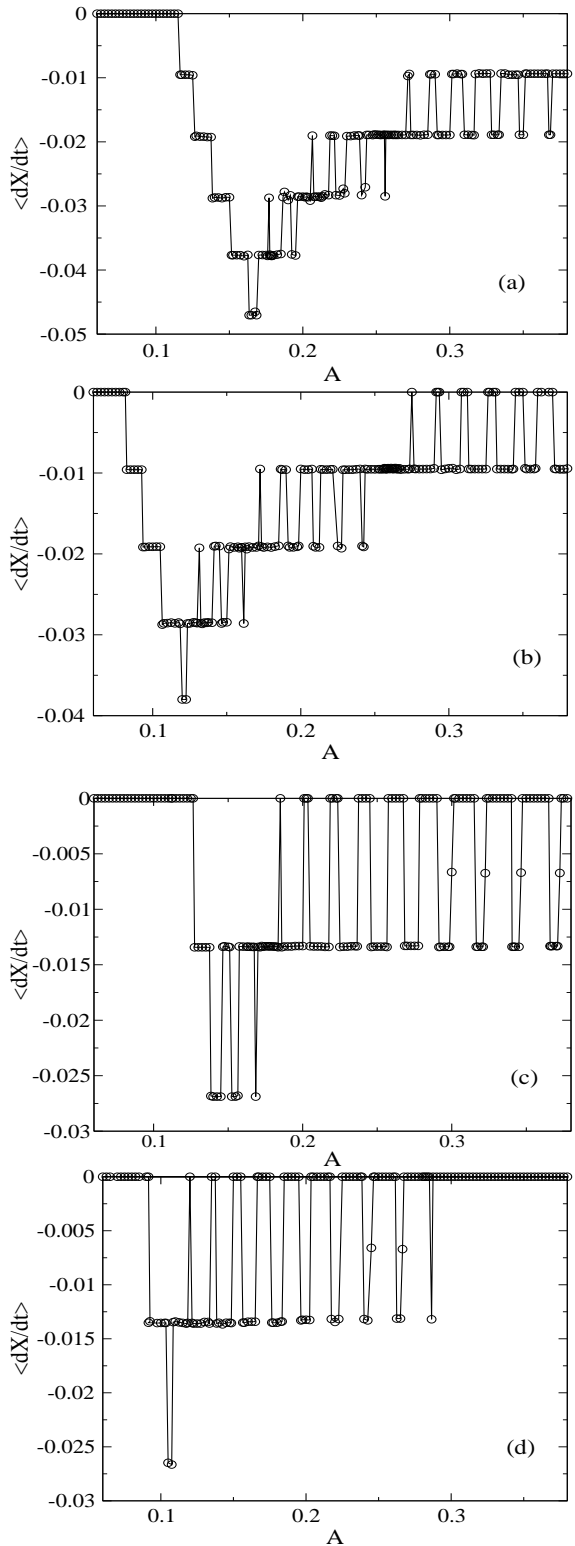

\begin{center}
\includegraphics[width=7.5cm,height=5cm]{dataomega=0.015e=0.8phi4-0.5-1.0-2.3L=4.eps}
\includegraphics[width=7.5cm,height=5cm,origin]{dataomega=0.015e=0.6phi4-0.5-1.0-2.3L=4.eps}
\\
\
\\
\includegraphics[width=7.5cm,height=5cm]{dataomega=0.015e=0.8phi4-0.7-1.4-3.2L=5.6.eps}
\includegraphics[width=7.5cm,height=5cm]{dataomega=0.015e=0.6phi4-0.7-1.4-3.2L=5.6.eps}
\end{center}
\caption{Simulation results for $\phi^4$: Mean velocity vs driving amplitude $A$ for the frequency $\omega=0.015$.(a) $\epsilon=0.8$, $x_{1}=0.5$, $x_{2}=1$, $x_{3}=2.3$, $L=4$.(b) $\epsilon=0.6$, $x_{1}=0.5$, $x_{2}=1$, $x_{3}=2.3$, $L=4$. (c) $\epsilon=0.8$, $x_{1}=0.7$, $x_{2}=1.4$, $x_{3}=3.2$, $L=5.6$. (d) $\epsilon=0.6$, $x_{1}=0.7$, $x_{2}=1.4$, $x_{3}=3.2$, $L=5.6$. The lines connecting the points serve as guides for the eye.}
\label{ratchet2}
\end{figure} 
It is clear from those expressions that, while 
in the case of $\phi^4$ we have a $\cosh^4$ factor in the denominator, 
sG has a $\cosh^2$ factor. Therefore, the peaks and valleys in the effective 
potential for the $\phi^4$ system are much narrower than for sG, thus 
compensating for the increment in length.
In addition, as in the sG model we will have dynamical changes of the
effective potential due to the kink width variations,
making more complicated the dynamics of motion.
In any event, 
these effective potentials obtained in the simple approach highlight the
importance of interference effects (see also \cite{angel,angel2}) and make it
clear that the kink width is not the unique quantity to take into account.

The consequences of choosing either the first or the second array for the
kink dynamics are revealed in Fig.\ \ref{ratchet2}. We have taken for the
analysis the $\phi^4$ model with a relative low frequency for the ac force,
for which the mean velocity as a function of the ac force amplitude shows a
staircase structure.
The range for the amplitude values was taken from the following rescaling
expression: $q^{\phi^4}A^{\phi^4}/M_{0}^{\phi^4}=q^{sG}A^{sG}/M_{0}^{sG}$.
This rule is deduced from the comparison between the 1-CC approaches for sG
and $\phi^4$ models, considering the similitude of the normalized potentials
discussed above.
Fig.\ \ref{ratchet2} shows the dependence of the kink mean
velocity as a function of the amplitude for two different arrays and
heights of the perturbations introduced by the inhomogeneities. 
As we can see, the motion is quantized as in standard ratchet systems
\cite{sironis,hangi} and is characterized by the existence of gaps for which
the net motion is absent (i.e., pure oscillating states). 
The mean velocity can be expressed as
$\vert\langle dX/dt \rangle\vert\equiv\vert\langle V
\rangle\vert=\frac{\textstyle L\omega}{\textstyle 2\pi}
\frac{\textstyle m}{\textstyle n}$ as usual \cite{barbi}, where the indices $m$,
$n$ $ \in \Bbb{N}$ quantize the motion. Using the expression for $\vert\langle
V \rangle\vert$ we can characterize the motion for each frequency and period
of the array.
Comparing the values obtained from the simulations with the results derived
from the expression for  $\vert\langle V \rangle\vert$ with corresponding 
parameters $L$ and $\omega$, we find that $m$ and $n$ can take
the following values: For panel a), $m=1,2,3,4,5$ and $n=1$; for panel b),
  $m=1,2,3,4$ and $n=1$; for panel c), $m=1,2$ with $n=1$ and $n=2$, and for 
panel d), $m=1,2$ with $n=1$ and $n=2$ .
Although the mean velocity increases with the spatial period, 
the maximum value of the index $m$ significantly decreases, leading to a global decreasing of the
velocity. 
These results prove that the inclusion of more inhomogeneities per cycle,
which obviously enhances the spatial period, is not a good option if we want
to reach high velocities. Furthermore, a very low frequency would be required
to obtain windows of motion.
In the case of the dependence on the inhomogeneities height, the starting point of the stair-steps structure  shows
a shift towards greater amplitude of the ac force when increases the height, which is natural in order to
overcome the barrier. Nevertheless, a higher speed is found, arising from a
higher $m$ and observable also as a broadening in the windows of motion. 
 
\section{Dynamics under noise}

This far, 
we have analyzed the ratchet-like behavior of our system in the 
deterministic case. However, it is clear that for our model to be more realistic,
for instance, in the context of LJJ, 
the effect of the temperature has to be taken into account. 
The behavior of ratchet systems for nonzero temperature has been extensively studied
both for point particles \cite{mag,doering,millonas,hangi,hangi2} and for
nonlinear extended systems \cite{extended,falo2,derenyi}. As for the problem 
we are considering here, the fact that there has not been much effort on 
soliton ratchet-like phenomena induced by spatial inhomogeneities carries over
to the stochastic effects. Therefore, it is important that we address this 
issue here. For the present work, we will focus on 
the robustness of our rocking ratchet under thermal fluctuations. Another 
relevant issue would be the possible activation, resonances or modifications of the 
transport features induced by noise, but this topic deserves a detailed 
analysis and will be the subject of future work. 

\subsection{Full model} 

For the sake of definiteness, 
we consider the sG model under the influence of a Gaussian white noise; the 
results for the $\phi^4$ equation are similar. 
Introducing the effect of the temperature through the fluctuation-dissipation
relationship and considering the overdamped case as before,
taking $\beta=1$, we find the following equation:
\begin{equation}
\ \phi_{tt}+\phi_{t}-\phi_{xx}+ \sin(\phi)[1+V(x)]=f(t)+ \eta(x,t), 
\end{equation}
with
\begin{eqnarray}
\langle\eta(x,t)\rangle&=&0,\nonumber\\
 \langle\eta(x,t)\eta(x',t')\rangle&=&D\delta{(x-x')}\delta{(t-t')}.
\end{eqnarray} 
\begin{figure}
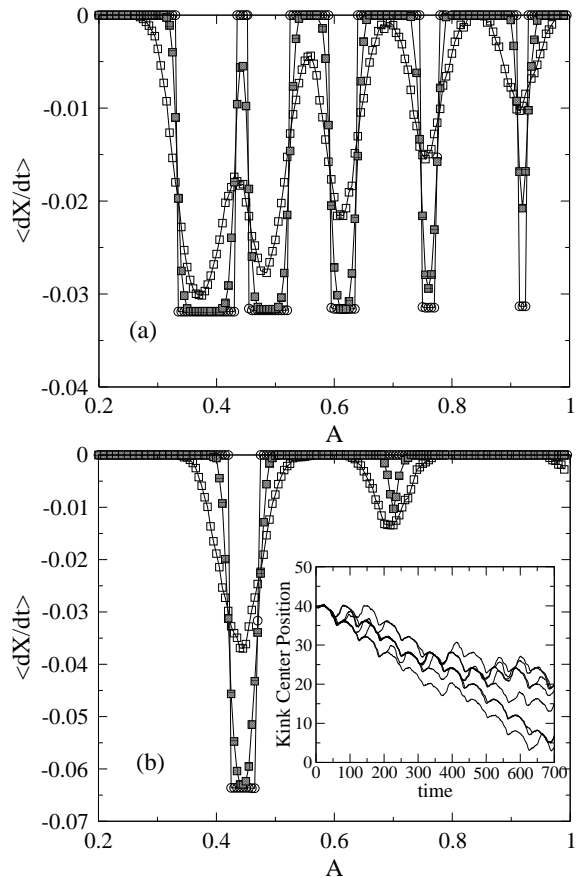

\begin{center} 
\includegraphics[width=7.5cm]{data1-D=0.-D=0.005-D=0.05-omega0.05.eps}
\includegraphics[width=7.5cm]{data1-D=0.-D=0.05-D=0.005-omega0.1+grafico-Kink-center-vs-time-8iter.eps}
\end{center}
\caption{Mean kink velocity $\langle dX/dt\rangle$ vs driving amplitude $A$ for
different intensities of the noise. (a) $\omega=0.05$: circles $D=0.$,
filled squares $D=0.005$, squares $D=0.05$; (b) $\omega=0.1$: circles $D=0.$,
filled squares $D=0.005$, squares $D=0.05$. Lines
serve as guides for the eye.
The inset shows several realizations for the motion of the kink center
with $A=0.43$, $\delta_{0}=\pi$ and $D=0.05$. }
\label{ratchet3}
\end{figure}
where $ f(t)\equiv A\sin(\omega t+\delta_{0})$ and the
intensity of the noise $D=2 k_{B}T$.

For the numerical simulations of the full partial differential equation as well 
as 
for the numerical solution of the collective variables (to be discussed in 
the next subsection), we have used
the Heun method with the Box-Muller-Wiener algorithm for generating
Gaussian random numbers of mean zero and variance one \cite{heun}.
In Fig.\ \ref{ratchet3} we show the behavior of the kink center dynamics
under thermal fluctuations. Hereafter, we have set the array parameters to be 
$x_{1}=0.5$, $x_{2}=1.$, $x_{3}=2.3$, $L=4$ and $\epsilon=0.8$ for our study.
The mean velocity was calculated using the expression in \cite{reiman}, namely
\begin{equation}\label{16}
\langle V\rangle=\langle \dot{X}\rangle=\lim_{t\rightarrow \infty}\frac{\langle X(t)-X(0) \rangle}{t},
\end{equation}
where the average is to be understood over many realizations of the noise.
From this figure we see that the steps of the deterministic case are now 
smoothed, a typical feature for the dynamics under noise.
It is important to realize that this smoothing affects the regions between the
windows, which become minima of the mean velocity modulus $\vert\langle V \rangle\vert $ instead of gaps with null mean velocity (see Fig.\
\ref{ratchet3}a). This phenomenon is directly related to the strength of the noise,
i.e, when the noise increases, the absolute value for the mean velocity decreases but simultaneously
the connection between the windows becomes more evident and the windows of
motion become less pronounced.

As in most other ratchet systems, in our model the stochastic fluctuations due
to temperature assist the jumps of the kink center from one well to the next one,
allowing in some cases jumps in the opposite direction of the rectification 
(see the inset graphic in Fig.\ \ref{ratchet3}b) which is not possible in the
absence of noise. Accordingly, the thermal fluctuations affect the mechanism
of rectification whereas, on the other hand, they yield the dynamics of the
pure oscillating states of the kink center unstable (i.e., destabilizing the
regions with locked directional motion for zero temperature). The joint action 
of both effects leads to the smoothing of the windows and the connection of 
the deterministic gaps.
\begin{figure}
\begin{center} 
\includegraphics[width=8cm]{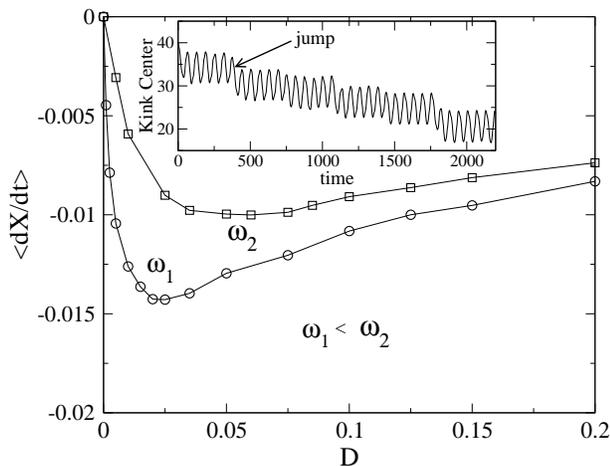}
\end{center}
\caption{Mean kink velocity $\langle dX/dt\rangle$ vs intensity of noise $D$.
Circles: $\omega=0.1$ and $A=0.70$; squares: $\omega=0.11$ and $A=0.75$.
Inset shows one realization for the motion of the kink center for
$\omega=0.1$, $A=0.70$, $\delta_{0}=0$ and $D=0.005$.}
\label{activa}
\end{figure}
For relative high temperatures the thermal kink energy is sufficient for
overcoming the barrier of the effective potential, and the kink motion is 
in practice diffusive, the influence of the barrier becoming negligible.
For this reason
the rectification of motion takes place only for not so large values of
the noise intensity (see discussion in \cite{sofia}).

A remarkable feature we have observed in the simulations
is shown in Fig.\ \ref{ratchet3}b for frequency $\omega=0.1$,
where new windows (absent in the deterministic case) appear.
This scenario is very similar to the one reported in \cite{falo2}
where a similar surprising and intriguing phenomenon was noted.
There, the authors discussed that these new windows arose due to 
jumps of the fluxons between stable and unstable pinned fixed points 
of the deterministic dynamics. Considering the interest for this 
 stochastic phenomenon, 
we carried out a careful analysis of the corresponding zone.
To summarize this work, in Fig.\ \ref{activa} we plot
the mean velocity as a function of the noise intensity
for different values of the frequency, showing the existence of an effective
value for the intensity of the noise for which a maximum absolute value of
the mean velocity is obtained.
The inset in Fig.\ \ref{activa} makes it clear that, as expected and suggested
in \cite{falo2}, 
the mechanism of activation occurs through jumps between multistable states 
(states of the kink center which in absence of noise are purely oscillating). 
Therefore, a higher velocity is obtained when the residence time in these
multistates is reduced or, in other words, when the intervals between
consecutive jumps decrease. Once again, this process of activation becomes
more effective when the noise intensity increases, but above a certain value
of the noise intensity the kink center starts to jump in the direction opposite
to that of the rectification, leading to a global loss in efficiency.
This explains the existence of an effective value for the noise intensity
for which the velocity reaches a maximum value. 

Another interesting characteristic observed in  Fig.\ \ref{activa} is
the dependence of the maximum mean velocity on the frequency. Specifically,
for a frequency value slightly larger than $\omega=0.1$, the maximum velocity
decreases, the peak moving towards greater values of the noise and the 
corresponding window of motion moving towards greater values of the ac force.
Accordingly, for a relatively large value of the frequency, above $\omega=0.11$,
the window of motion induced by noise disappears.
On the other hand, for frequencies slightly smaller than $\omega=0.1$,
a new window in absence of noise is obtained. 
With all these results, it is clearly established that the unidirectional motion
induced by noise occurs only for a narrow window of frequency values. 

We will show in the next subsection that this phenomenon seems to be a general
feature, since at the CC level the system behaves very
much like the dynamics of point particles.

\subsection{Collective Coordinates in presence of noise}

In order to understand the behavior observed in the previous section we 
resort again to the CC approach. 
As a first step, we take only into account one degree of freedom.
Although, as discussed above, this framework is inaccurate for describing
quantitatively the kink motion on a lattice of inhomogeneities,
it does help understand qualitatively most of the features observed in the
simulations, without unnecessary analytical complications. 
After some algebra (see Appendix A for details), with $\beta=1$, we find
the following expression for the kink center coordinate with noise:
\begin{equation}\label{17}
M_{0}\ddot{X}+ M_{0}\dot{X}=-\frac{d\
U}{d X}-q f(t)+\sqrt{D M_{0}}\,\ \xi(t)
\end{equation}
with $\langle\xi(t)\rangle=0$, $\langle\xi(t)\xi(t')\rangle=\delta(t-t')$.
For the sake of simplicity we have taken the nonrelativistic approach
$\dot{X}^2\ll 1$,
for which the noise turns out to be simply additive. 

\begin{figure}
\begin{center} 
\includegraphics[width=8cm]{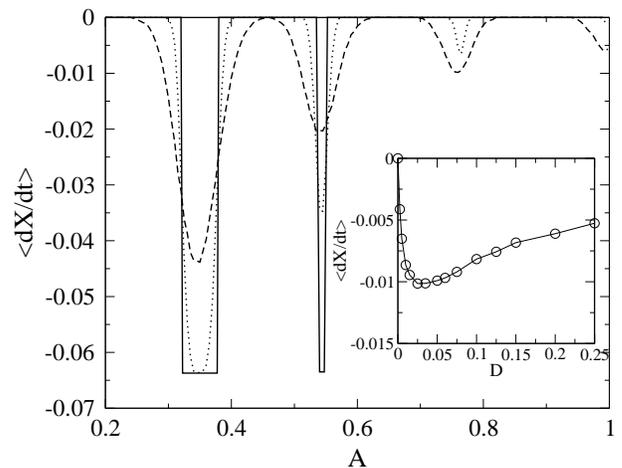}
\end{center}
\caption{CC approach: Mean kink velocity  
$\langle dX/dt\rangle$ vs driving amplitude $A$
for different intensities of the noise and frequency $\omega=0.1$. Continuous line: $D=0$, dotted line:
$D=0.005$, dashed line: $D=0.05$. Inset: Mean kink velocity
$\langle dX/dt\rangle$ vs intensity of noise $D$ for $A=0.7625$.}
\label{ratchet4}
\end{figure}
Figure \ref{ratchet4} presents the results of the numerical integration of
Eq.\ (\ref{17}). Much as we did in the simulations, we calculate the mean
velocity using Eq.\ (\ref{16}), taking up to 500 realizations.
From this plot two main features also observed in the simulations can be seen.
First, smooth curves are obtained for the mean velocity as a function of the
amplitude of the ac force, with values that decrease when the noise is increased.
Second, new windows appear, and inside them there is a value of the noise intensity
for which the velocity reaches a maximum value (inset in Fig.\ 9). 
It is thus evident that,
in spite of the quantitative differences with the simulations,
this simple approach does predict correctly the qualitative behavior of the full 
system.

In order to improve the results presented so far, we have extended the
framework to two collective variables. By doing so (see Appendix B) 
we arrive at Eqs.\ (\ref{B21})-(\ref{B22}) with two uncorrelated {\it multiplicative}
white noises, meaning that the noises depend on the kink width dynamics.
\begin{figure}
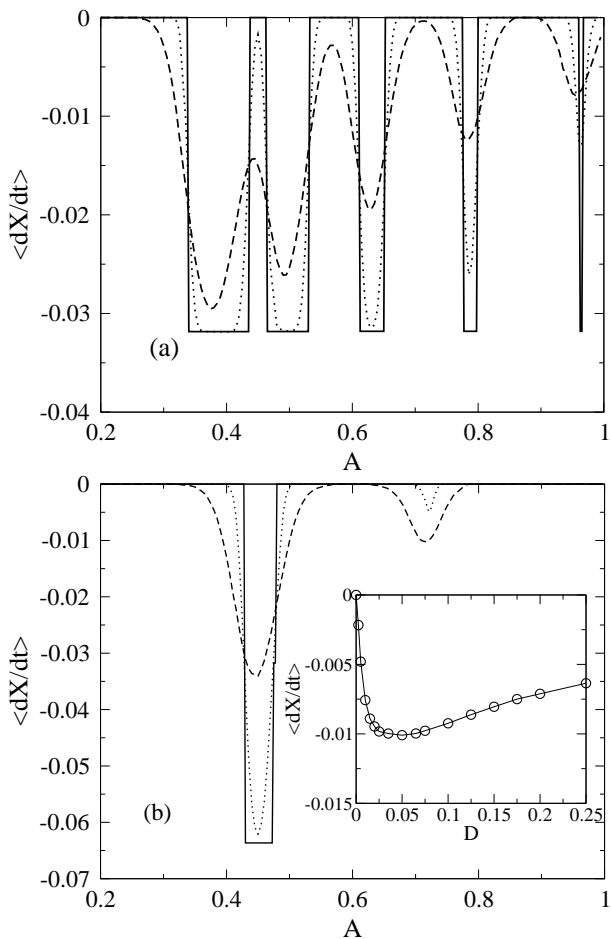

\begin{center}
\includegraphics[width=8.cm]{cc3omega=0.05-D=0.-D=0.005-D=0.05-500iter.eps} 
\includegraphics[width=8.cm]{cc3omega=0.1-D=0.-D=0.05-D=0.005-500iter+inset.eps}
\end{center}
\caption{CC Approach for two degrees of freedom (Eqs.(\ref{B21})-(\ref{B22}) of the appendix). Mean kink velocity $\langle dX/dt\rangle$ vs driving amplitude $A$ for different intensities of the noise. (a) $\omega=0.05$; continuous line: $D=0$, dotted line: $D=0.005$, dashed line: $D=0.05$; (b) $\omega=0.1$; continuous line: $D=0$, dotted line: $D=0.005$, dashed line: $D=0.05$.  Inset shows Mean kink velocity $\langle dX/dt\rangle$ vs Intensity of Noise $D$ for $A=0.72$ with $\omega=0.1$.}
\label{ratchet5}
\end{figure}
The results for this improved approach are collected in  Fig.\ \ref{ratchet5}.
Comparing with the simulations (Fig.\ \ref{ratchet3}), we can observe the
excellent agreement, with the locations of the windows correctly predicted.
As expected the curves are again smooth, a feature correctly accounted for
already in the 1-CC framework.
For the frequency $ \omega=0.1$, a new window is predicted, whose location is also in very
good agreement with its corresponding window in the simulations.
These results confirm the importance of considering the kink width dynamics
in the framework of the collective coordinates in order to achieve correct quantitative
results as compared to the simulations.
  
\section{Conclusions} 

We have studied the dynamics of solitons in a lattice of point-like
inhomogeneities in two different nonlinear systems, the sG and the $\phi^4$ models,
as good representatives of an entire class of nonlinear models. We have designed a 
periodic lattice made of an asymmetric unit cell, where, depending on the lattice
parameters, ratchet-like motion of solitons (rectification) is observed.
This is one
of the few examples proposed so far for soliton ratchets based on spatial 
inhomogeneity.
Building on
a preliminary work reported in \cite{pub}, we have analyzed in full 
detail all the characteristics and features of interest of our model.
In order to understand the observed phenomenology, we have 
developed two CC approaches: one that takes into account only the motion of
the kink center, and another one that includes the kink width as a second 
degree of freedom. We have seen that whereas both techniques give good
qualitative results and allow for a general understanding the phenomenon, 
only when the kink width enters in the CC description the results are 
quantitatively correct. 
Interestingly, the CC approach allows to show that rectification takes place 
only when the unit cell gives rise to an effective potential of the same 
length scale as the kink width. This fact, along with the relevant role 
played by the width dynamical behavior, makes us conclude that one essential
ingredient for observing ratchet-like behavior in this system is the
existence of length scale competition \cite{angel0}. 

As a necessary complement of the deterministic study, we have considered the 
influence of thermal noise on the behavior of our model systems. As in the 
deterministic case, the CC technique yields very good results. Of particular
relevance is the appearance, at the 2-CC level, of multiplicative noise in
the CC equations, coupled to the kink width. This is yet another hint to the
crucial role played by this degree of freedom in the dynamics of the system. 
Another important feature is the motion activation induced by noise, that
was observed in the simulations and later corroborated in the CC framework.
 
In a wider context, we believe that the CC approach presented here, and the
conclusions about the role of the kink width, are of a much more general 
applicability. This is the case, for instance, with the anomalous 
resonances observed in nonlinear Klein-Gordon models \cite{niurka2}, which
can only be explained by considering the width degree of freedom. 
Specifically within the context of rectification,
we have previously shown that the existence of ratchet behavior
induced by pure asymmetric temporal driving in a homogeneous system, 
recently demonstrated in experiments in LJJ \cite{Ustinov}, is 
due to the coupling between the translation and the width oscillations \cite{Niur}. 
Based on all these experiences, we believe that 
the implementation of the 2-CC approach can be very useful in the case
of spatially correlated noise \cite{marchesoni3,braun}, where the 1-CC approach 
can not explain the length scale competition between the correlation length and
the size of the kink, first suggested in \cite{marchesoni3}. The new formulation
would be also necessary in using the CC framework for describing the transport of
proteins assisted by a thermal bath provoked by ATP molecules hydrolization,
where the width is an important quantity to take into account. 
On the other hand, it is becoming more and more evident the crucial 
contribution of internal degrees of freedom in the functioning of 
molecular motors \cite{kinesin}. In this respect, our results suggests
that models including in a natural way this degree of freedom, like 
in our case, can be the proper descriptions of those phenomena. 
Size dependent rectification has been also reported in colloids 
\cite{marquet} and rectification through clustering has been 
observed in granular gases \cite{meer}, which further reinforce our
conclusion that the deformation or internal degrees of freedom must
be an ingredient of good theoretical description of ratchet 
phenomena beyond the point particle scenario.

Finally, we want to stress that our design of a soliton system with 
ratchet behavior is a very simple one, that can be implemented in actual 
experiments and devices such as LJJ, for instance. Another field where
the current state-of-the-art allows to apply this result is that of 
engineered molecular motors, demonstrated, e.g., in \cite{tsi}. 
In this case, our proposal may apply to the design of biomolecular
devices with medical applications. Generally speaking, 
this type of approach to rectification can be of 
interest for applications in which it is needed to have a tunable 
rectifier tailored for a specific regime. Of course, experimental 
verification of our predictions is needed to ascertain the accuracy of
our results. We hope that this work stimulates experiments in this 
direction. Indeed the research reported here, 
opens new perspectives in the design of ratchet devices 
for more complicated extended nonlinear systems, such as general 
coupled chains \cite{kivshar2}. Of particular interest in this class 
are stacked LJJ \cite{kivshar3}, although there are many other systems 
with potential applications in different areas. 
Work along this line is in progress.

\acknowledgments 

This work has been
supported by the Ministerio de Ciencia y Tecnolog\'\i a of Spain
through grant BFM2003-07749-C05-01 (AS)
and by the International Research Training Group
`Nonequilibrium Phenomena and Phase Transitions in Complex Systems'
(DFG, Germany). The authors also want to thanks the comments of N.R. Quintero.

\appendix
 
\section{Collective Coordinates. First approach}

In order to apply the GTWA first proposed in \cite{mertens} we rewrite Eq.(\ref{1}) adding Gaussian white noise 

\begin{eqnarray} \label{A1}
 \dot{\phi}&=&\frac{\delta H}{\delta \psi},\>\>\> 
\\
\dot{\psi}&=&-\frac{\delta H}{\delta \phi}-\beta\dot{\phi}-\frac{\partial \widetilde{U}}{\partial \phi}V(x)+f(t)+\eta(x,t) \label{A2}
\end{eqnarray} 

with 

\begin{eqnarray}\label{A3}
\langle\eta(x,t)\rangle&=&0,\nonumber\\
\\
 \langle\eta(x,t)\eta(x',t')\rangle&=&D\delta{(x-x')}\delta{(t-t')},\nonumber
\end{eqnarray}

where $\psi= \dot{\phi}$, $ f(t)\equiv A\sin(\omega t+\delta_{0})$, $D=2\beta k_{B}T$ and $H$ is the Hamiltonian corresponding to the unperturbed form of Eq.(\ref{1}) given by
\begin{equation}\label{A4}
\begin{array}{l}
{\displaystyle{ 
{\it{H}} = \int_{-\infty}^{+\infty} dx \Big \{ 
\frac{1}{2} \psi^2 + \frac{1}{2} \phi_{x}^{2} + U(\phi) \Big\}}}.
\end{array}
\end{equation}

As starting point we assume that the solution has the form
 \begin{equation}\label{A5}
\phi(x,t)=\phi_{K}[x-X(t),\dot{X}],
\end{equation}

and therefore by definition of $\psi$ we have that
 
 \begin{equation} \label{A6}
\psi(x,t)=\psi_{K}[x-X(t),\dot{X}(t),\ddot{X}].
\end{equation}

The index $K$ refers to the kink shape, but in the following we will omit it for simplicity.

Following the procedure in \cite{mertens}, inserting $\dot{\phi}$, $\dot{\psi}$ into Eqs.(\ref{A1})-(\ref{A2}) we get 
the expressions

\begin{equation}\label{A7}
\frac{\partial \phi}{\partial X}\dot{X}+\frac{\partial \phi}{\partial \dot{X}}\ddot{X}=\frac{\delta H}{\delta \psi},\qquad  
\end{equation}

\begin{eqnarray}
\frac{\partial \psi}{\partial X}\dot{X}+\frac{\partial \psi}{\partial \dot{X}}\ddot{X}+\frac{\partial \psi}{\partial \ddot{X}}\dddot{X}=-\frac{\delta H}{\delta \phi}-\beta\left(\frac{\partial \phi}{\partial X}\dot{X}+\frac{\partial \phi}{\partial\dot{X}}\ddot{X}\right) \nonumber \\
-\frac{\partial\widetilde{U}}{\partial \phi}V(x)+f(t)+\eta(x,t).\hspace{1cm}  \label{A8}
\end{eqnarray}

Multiplying Eq.(\ref{A7}) by $\frac{\textstyle \partial \psi}{\textstyle \partial X}$ and Eq.(\ref{A8}) by $\frac{\textstyle \partial \phi}{\textstyle \partial X}$, and then subtracting both expressions and integrating we arrive at the following equation

\begin{equation}
\  A \dddot{X}+ M\ddot{X}=-\beta C_1\dot{X}-\beta C_2\ddot{X} +F^{ac}+F^{stat}+F^{inh}+F^{st},\label{A9}
\end{equation}

whose values for the coefficients and forces are given by 
\begin{eqnarray}
\lefteqn{A=\int_{-\infty}^{\infty} dx \frac{\partial \phi}{\partial X} \frac{\partial \psi}{\partial \ddot{X}},\qquad F^{ac}=\int_{-\infty}^{\infty} dx\,\ f(t)   \frac{\partial \phi}{\partial X},}\nonumber \\ \nonumber
\lefteqn{ C_1=\int_{-\infty}^{\infty} dx\left(\frac{\partial\phi}{\partial X}\right)^2, \,\,\
F^{inh}=-\int_{-\infty}^{\infty} dx \frac{\partial \widetilde{U}}{\partial \phi} V(x)\frac{\partial \phi}{\partial X},} \\ \nonumber
&&C_2=\int_{-\infty}^{\infty} dx\frac{\partial\phi}{\partial X}\frac{\partial\phi}{\partial\dot{X}},\qquad
F^{st}=\int_{-\infty}^{\infty} dx \,\ \eta(x,t)\frac{\partial \phi}{\partial X}, \\ \nonumber 
\lefteqn{M=\int_{-\infty}^{\infty} dx\left( \frac{\partial \psi}{\partial \dot{X}}\frac{\partial \phi}{\partial X}- \frac{\partial \phi}{\partial \dot{X}} \frac{\partial \psi}{\partial X} \right),}  \\  \nonumber
\lefteqn{F^{stat} = -\int_{-\infty}^{+\infty} dx \,\ \left \{  
\frac{\delta {\it{H}}}{\delta \phi} \frac{\partial \phi}{\partial X}  +  
\frac{\delta {\it{H}}}{\delta \psi} \frac{\partial \psi}{\partial X} 
\right \}} \nonumber 
\\ 
&&=-\int_{-\infty}^{+\infty} dx \,\ \frac{\partial {\cal{H}}}{\partial X} = 
- \frac{\partial E}{\partial X},\nonumber
\end{eqnarray}
where $E$ represents the energy of the system, $\cal{H}$ is the Hamiltonian density of Eq.(\ref{A4}) and $F^{stat}$ is the static force due to the external field, equal to zero for the above Hamiltonian. 

Next we consider the sG potential for the system Eqs.(\ref{A1})-(\ref{A2}) for which we assume as solution the ansatz 
\begin{equation} \label{A14}
\phi(x,t)=\phi^{(0)} [\gamma(x-X(t))]=4\arctan\left(\exp\left[\gamma(x-X(t))\right]\right),
\end{equation}
where  $\phi^{(0)}=4\arctan\left(\exp\left[(x-X_{0})/l_{0}\right]\right)$ is the static kink solution of the sG system, centered in $X_{0}$ and of width $l_{0}$. Here  $\gamma=1/\sqrt{1-\dot{X}^2}$ where we have put $l_{0}=1$ for the sG case.

 Considering the previous statement for the static force and taking into account $V(x)$ from Eq.(\ref{2}), we obtain

\begin{eqnarray}
A&=&0, \,\ \qquad  F^{ac}=-q f(t),\nonumber \\
M&=&\gamma^3 M_{0}, \qquad    F^{stat}= 0,\nonumber \\
C_1&=&\gamma M_{0},\qquad F^{inh}=- \frac{\partial U}{\partial X}, \nonumber \\
C_2&=&0,\nonumber
\end{eqnarray}

where $M_{0}=8$ is the kink mass, $q=2\pi$ is the topological charge and $U(X,\dot{X})$ given by
\begin{equation}
U(X,\dot{X})=2\epsilon \sum_{n}\sum_{i=1}^{3}\frac{1}{\cosh^2[\gamma(X-x_{i}-nL)]}\end{equation}
is the effective potential. In the non-relativistic limit $\dot{X}^2\ll 1$, $U(X,\dot{X})\simeq U(X)$.
                    
A representation for the stochastic force $F^{st}$ can be obtained from the calculation of the variance. In the case of additive noise it is allowed to make the following assumption

\begin{eqnarray}
\lefteqn{\langle \frac{\partial \phi^{(0)}(x,t)}{\partial X}\frac{\partial \phi^{(0)}(x',t')}{\partial X}\,\ \eta(x,t) \eta(x',t') \rangle}  \nonumber  \\
&=&\frac{\partial \phi^{(0)}(x,t)}{\partial X}\frac{\partial \phi^{(0)}(x',t')}{\partial X}\,\ \langle\eta(x,t) \eta(x',t') \rangle. \label{A12}
\end{eqnarray}

Hence the correlation function for $F^{st}$ can be written as

\begin{eqnarray}
\lefteqn{\langle F^{st}(t) F^{st}(t')\rangle}  \nonumber  \\
&=&\int_{-\infty}^{\infty}\int_{-\infty}^{\infty}dx dx' \frac{\partial \phi^{(0)}(x,t)}{\partial X}\frac{\partial \phi^{(0)}(x',t')}{\partial X} \langle\eta(x,t) \eta(x',t') \rangle,\nonumber \\
\end{eqnarray}
for which, taking into account the expression (\ref{A2}), after some algebra we get

\begin{equation}
\langle F^{st}(t) F^{st}(t')\rangle=2\beta k_{B}T\gamma M_{0}\delta(t-t'),
\end{equation}

i.e., $ F^{st}(t)$ is a white noise with kink diffusion constant 
$$D_{K}=\gamma M_{0} D.$$ 

As a consequence we obtain a non-additive noise due to the factor $\gamma(\dot{X})$, i.e, we have a problem with multiplicative noise.

Then the equation of motion (\ref{A9}) can be rewritten as
\begin{equation}
\gamma^3M_{0}\ddot{X}+\beta\gamma M_{0}\dot{X}=-qf(t)- \frac{\partial U}{\partial X}+
\sqrt{D_{K}}\,\ \xi(t) \label{A15}
\end{equation}

with $\langle\xi(t)\rangle=0$, $\langle\xi(t)\xi(t')\rangle=\delta(t-t')$.
The Eq.(\ref{A15}) in absence of inhomogeneities and noise agrees with the results presented in \cite{niurka3}. 
The other r.h.s. terms that appear in (\ref{A15}) are in correspondence with those already obtained in \cite{angel,joergensen} in the presence of impurities (non-relativistic approach) and Gaussian white noise, respectively. The procedure used here is equivalent to the so-called adiabatic approach by using modified conservation laws \cite{malomed}.

\section{Collective Coordinates. Second approach}
In order to get the CC equations we follow a similar procedure as in the previous section but this time we propose a solution with the form 

\begin{equation}\label{B1}
\phi(x,t)=\phi[x-X(t),l(t)],
\end{equation}

 \begin{equation} \label{B2}
\psi(x,t)=\psi[x-X(t),l(t),\dot{X}(t),\dot{l}]
\end{equation}
with $\psi=\dot{\phi}$, 
which considers the kink width as a new collective variable
(see e.g. \cite{niurka2}).

Inserting Eqs.(\ref{B1}) and (\ref{B2}) in our system Eqs.(\ref{A1})-(\ref{A2}) and then multiplying the first equation by  $\frac{\textstyle \partial \psi}{\textstyle \partial X}$ and 
the second one by $\frac{\textstyle \partial \phi}{\textstyle \partial X}$; subtracting both expression and integrating we arrive at the following equation

\begin{eqnarray}\label{B3}
&&\int_{-\infty}^{+\infty} dx \frac{\partial \phi}{\partial X} 
\frac{\partial \psi}{\partial \dot{X}} \ddot{X} +   
\int_{-\infty}^{+\infty} dx   [\phi, \psi]  \dot{l} + 
\int_{-\infty}^{+\infty} dx  \frac{\partial \phi}{\partial X} 
\frac{\partial \psi}{\partial \dot{l}}  \ddot{l} 
\nonumber 
\\
&&- F^{stat} = \,\ \int_{-\infty}^{+\infty} dx\,\ F(x,t,\phi,\phi_{t},...) \frac{\partial \phi}{\partial X}\qquad 
\end{eqnarray}

with $F(x,t,\phi,\phi_{t},...)=-\beta\dot{\phi}-\frac{\textstyle \partial \widetilde{U}}{\textstyle \partial \phi}V(x)+f(t)+\eta(x,t)$, and

\begin{equation}
[\phi, \psi] =   
 \frac{\partial \phi}{\partial X} \frac{\partial \psi}{\partial l} - 
 \frac{\partial \phi}{\partial l} \frac{\partial \psi}{\partial X},
\end{equation}

\begin{eqnarray} 
F^{stat}& =& -\int_{-\infty}^{+\infty} dx \,\ \left \{  
\frac{\delta {\it{H}}}{\delta \phi} \frac{\partial \phi}{\partial X}  +  
\frac{\delta {\it{H}}}{\delta \psi} \frac{\partial \psi}{\partial X} 
\right \} \nonumber 
\\ 
&=&-\int_{-\infty}^{+\infty} dx \,\ \frac{\partial {\cal{H}}}{\partial X}, 
\end{eqnarray}

where $\cal{H}$ is the Hamiltonian density of Eq.(\ref{A4}) for which, as was seen before, a null value for $F^{stat}$ is obtained.  

Repeating the same procedure, but now with $\frac{\textstyle \partial \psi}{\textstyle \partial l}$ and  $\frac{\textstyle \partial \phi}{\textstyle \partial l}$, we get the expression

\begin{eqnarray}
&&\int_{-\infty}^{+\infty} dx  [\psi, \phi]  \dot{X}  + 
\int_{-\infty}^{+\infty} dx  \frac{\partial \phi}{\partial l} 
\frac{\partial \psi}{\partial \dot{X}} \ddot{X}+\int_{-\infty}^{+\infty} dx  \frac{\partial \phi}{\partial l} 
\frac{\partial \psi}{\partial \dot{l}}  \ddot{l} \nonumber
\\ 
&&
- K^{int} =  \int_{-\infty}^{+\infty} dx \,\ F(x,t,\phi,\phi_{t},...) 
\frac{\partial \phi}{\partial l}. \label{B6}
\end{eqnarray}

Following Rice \cite{rice} for the particular case of sG 

 \begin{equation} \label{B7}
\phi(x,t)=\phi^{(0)}[x-X(t),l(t)]=4\arctan\left(\exp\left[\frac{x-X(t)}{l(t)}\right]\right),
\end{equation}

Eq.(\ref{B3}) becomes

\begin{equation}\label{B8}
\ M_{0}l_{0}\frac{\ddot{X}}{l}+\beta M_{0}l_{0}\frac{\dot{X}}{l}-M_{0}l_{0}
\frac{\dot{X}\dot{l}}{l^2}=F^{ac}+F^{inh}+F^{st}
\end{equation}

with 

\begin{eqnarray}
F^{ac}&=&\int_{-\infty}^{\infty} dx\,\ f(t) \frac{\partial \phi^{(0)}}{\partial X}=-2\pi f(t)=-q f(t),
\\
F^{inh}&=&-\int_{-\infty}^{\infty} dx\,\ \sin(\phi^{(0)}) V(x) \frac{\partial \phi^{(0)}}{\partial X}=- \frac{\partial U}{\partial X},\qquad
\\
F^{st}&=&\int_{-\infty}^{\infty} dx\,\ \eta(x,t)\frac{\partial \phi^{(0)}}{\partial X},\label{B11}
\end{eqnarray}

 and 
\begin{equation}
U(X,l)=2\epsilon \sum_{n}\sum_{i=1}^{3}\frac{1}{\cosh^2[(X-x_{i}-nL)/l]}.
\end{equation}

On the other hand, Eq.(\ref{B6}) is transformed into 

\begin{equation}\label{B13}
\alpha
  M_{0}l_{0}\frac{\ddot{l}}{l}+\beta\alpha
  M_{0}l_{0}\frac{\dot{l}}{l}+M_{0}l_{0}\frac{\dot{X}^2}{l^2}=
K^{int}(l,\dot{l},\dot{X})+K^{inh}+K^{st}
\end{equation}

with 

\begin{eqnarray}
K^{inh}&=&-\int_{-\infty}^{\infty} dx\,\ \sin(\phi^{(0)}) V(x) \frac{\partial \phi^{(0)}}{\partial l}=- \frac{\partial U}{\partial l},\qquad
\\
K^{st}&=&\int_{-\infty}^{\infty}dx \,\ \eta(x,t)\frac{\partial \phi^{(0)}}{\partial l}, 
\end{eqnarray}

\begin{equation}
K^{int}(l,\dot{l},\dot{X})  = - \int_{-\infty}^{+\infty} dx  
\frac{\partial {\cal{H}}}{\partial l} = 
- \frac{\partial E}{\partial l}, 
\end{equation}

where $\alpha=\pi^2/12$, $M_{0}=8$, $l_{0}=1$ and

\begin{equation}
E = \frac{1}{2} \frac{l_{0}}{l} M_{0}  \dot{X}^{2} + 
\frac{1}{2} \frac{l_{0}}{l} \alpha M_{0}  \dot{l}^{2} + 
\frac{1}{2} M_{0} \left(\frac{l_{0}}{l} + \frac{l}{l_{0}}\right).
\end{equation}

As in the previous section we use the variances of the stochastic forces in order to obtain approximate expressions for it. Taking the assumption given by the expression (\ref{A12}) we find  
for (\ref{B11}) the correlation function
\begin{eqnarray}
\lefteqn{\langle F^{st}(t) F^{st}(t')\rangle} \nonumber  \\
&=&\int_{-\infty}^{\infty}\int_{-\infty}^{\infty}dx dx' \frac{\partial \phi^{(0)}(x,t)}{\partial X}\frac{\partial \phi^{(0)}(x',t')}{\partial X} \langle\eta(x,t) \eta(x',t') \rangle\nonumber
\\
&=&D\delta(t-t')\int_{-\infty}^{\infty}dx \left(\frac{\partial \phi^{(0)}}{\partial X}\right)^2=D\delta(t-t')\frac{l_{0}}{l}M_{0}.
\end{eqnarray}

In what follows  similar expressions to the Eq.(\ref{A12}) valid for additive noise are used in order to calculate other correlation functions like

\begin{eqnarray}
\lefteqn{\langle K^{st}(t) K^{st}(t')\rangle}  \nonumber  \\
&=&\int_{-\infty}^{\infty}\int_{-\infty}^{\infty}dx dx' \frac{\partial \phi^{(0)}(x,t)}{\partial l}\frac{\partial \phi^{(0)}(x',t')}{\partial l} \langle\eta(x,t) \eta(x',t') \rangle \nonumber
\\
 &=&D\delta(t-t')\int_{-\infty}^{\infty}dx \left(\frac{\partial \phi^{(0)}}{\partial l}\right)^2=D\delta(t-t')\frac{l_{0}}{l}\alpha M_{0}, \nonumber \\
\end{eqnarray}

and 

\begin{eqnarray}
\lefteqn{\langle F^{st}(t) K^{st}(t')\rangle} \nonumber  \\
&=&\int_{-\infty}^{\infty}\int_{-\infty}^{\infty}dx dx' \frac{\partial \phi^{(0)}(x,t)}{\partial X}\frac{\partial \phi^{(0)}(x',t')}{\partial l} \langle\eta(x,t) \eta(x',t') \rangle \nonumber
\\
&=&D\delta(t-t')\int_{-\infty}^{\infty}dx\,\ \frac{\partial \phi^{(0)}}{\partial X}\frac{\partial \phi^{(0)}}{\partial l}=0.
\end{eqnarray}
\

From the latter correlation for the stochastic forces we see that these are not cross-correlated. 

Finally, collecting all the previous results we can rewrite Eqs.(\ref{B8}), (\ref{B13}) as follows

\begin{eqnarray}
&&\ M_{0}l_{0}\frac{\ddot{X}}{l}+\beta M_{0}l_{0}\frac{\dot{X}}{l}-M_{0}l_{0}
\frac{\dot{X}\dot{l}}{l^2}=-\frac{\partial U}{\partial X}-qf(t)\nonumber \\
&& \hspace{4cm}
+\sqrt{\frac{D M_{0}l_{0}}{l}}\,\ \xi_{1}(t),\label{B21}\\
&&\alpha M_{0}l_{0}\frac{\ddot{l}}{l}+\beta\alpha
  M_{0}l_{0}\frac{\dot{l}}{l}+M_{0}l_{0}\frac{\dot{X}^2}{l^2}=-\frac{\partial U}{\partial l}+ K^{int}(l,\dot{l},\dot{X}) \nonumber \\
&&\hspace{4.cm}+\sqrt{\frac{D \alpha M_{0}l_{0}}{l}}\,\ \xi_{2}(t)\label{B22}
\end{eqnarray}

with
$\langle\xi_{i}(t)\rangle=0$,
$\langle\xi_{i}(t)\xi_{j}(t')\rangle=\delta_{ij}\delta(t-t')$, for $i,j=1,2$.\newline
A feature of particular interest in these new equations is the presence of stochastic forces which are multiplicative white noises dependent on the kink width variable.

The method described here using the technique of projection is equivalent to the variational calculations of the
momentum and the energy of the system for perturbed nonlinear Klein-Gordon systems of the form
of Eqs.(\ref{A1}) and (\ref{A2}) and with a Hamiltonian of the form of Eq.(\ref{A4}) (see \cite{niurka2} for details). Another procedure and derivation has been recently presented in \cite{elias}.

\end{document}